\documentclass[english]{article}
\usepackage[T1]{fontenc}
\usepackage[latin1]{inputenc}
\usepackage{setspace}
\usepackage{amssymb}

\makeatletter

\newcommand{\lyxaddress}[1]{
\par {\raggedright #1
\vspace{1.4em}
\noindent\par}
}

\usepackage{babel}
\makeatother
\begin{document}

\title{Generalized Electromagnetic Fields of Dyons in Isotropic Medium}

\author{Jivan Singh{*}, P. S. Bisht{*}{*} and O. P. S. Negi{*}{*}}

\maketitle

\lyxaddress{\begin{center}{*}Department of Physics\\
 M. B. Govt. P. G. College\\
 Haldwani(Nainital)\\
UA -263139, INDIA\par\end{center}}

\lyxaddress{\begin{center}{*}{*} Department of Physics\\
 Kumaun University\\
 S. S. J. Campus\\
 Almora-263601 (UA), INDIA\par\end{center}}

\lyxaddress{\begin{center}Email:-jgaria@indiatimes.com\\
 ~~~~~ps\_bisht123@rediffmail.com\\
 ~~~~~ops\_negi@yahoo.co.in\par\end{center}}

\begin{abstract}
Strating with the Maxwell's equations in presence of electric and
magnetic sources in an isotropic homogenous medium, we have derived
the various quantum equations of dyons in consistent and manifest
covariant way. It has been shown that the presented theory of dyons
remains invariant under the duality transformations in isotropic homogeneous
medium.
\end{abstract}
~~~~~~~Few interest in the subject of monopoles and dyons was
enhanced by the work of t' Hooft \cite{key-1} and Polyakov \cite{key-2}
and its extension by Julia and Zee \cite{key-3}. Consequently, these
particles became intrinsic part of all current grand unified theories
\cite{key-4,key-5} in view of their enormous potential importance
\cite{key-6,key-7,key-8,key-9}. Keeping in view the potential importance
of monopoles and the results of Witten \cite{key-9} that monopoles
are necessarily dyons, we \cite{key-10,key-11} have also constructed
a self-consistent co-variant theory of generalized electromagnetic
fields associated with dyons each carrying the generalized charge
as complex quantity with its real and imaginary part as electric and
magnetic constituents. On the other hand Kravchenko \cite{key-12}
has analysed the Maxwell's equations for time-dependent electromagnetic
fields in homogeneous (isotropic) and chiral medium. In this paper
we have derived the various quantum equations of generalized electrmagnetic
fields of dyons (particles carrying simultaneously electric and magnetic
charges) in isotropic medium in consistent and manifest co-variant
ways. It has been shown that the present theory of dyons remains invariant
under the duality transformations in isotropic homogeneous medium.
It has also been shown that the equation of motion in isotropic medium
reproduces the rotationally symmetric gauge invariant angular momentum
of dyons and accordingly we have derived the quantization condition
for generalised electromagnetic fields of dyons in isotropic medium.

Let us start with the symmetrized Maxwell's equations, derived by
Dirac \cite{key-13} in presence of magnetic charge (monopole) to
establish the dual invariance between electric and magnetic fields,
in the following manner in vacuum in SI units \cite{key-14} for $c=\hbar=1$i.e.\begin{eqnarray}
\overrightarrow{\nabla}\cdot\overrightarrow{D} & = & \rho_{e}\label{eq:1}\\
\overrightarrow{\nabla}\cdot\overrightarrow{B} & = & \mu_{0}\rho_{m}\label{eq:2}\\
\overrightarrow{\nabla}\times\overrightarrow{E} & = & \frac{\overrightarrow{j_{m}}}{\epsilon_{0}}-\frac{\partial\overrightarrow{B}}{\partial t}\label{eq:3}\\
\overrightarrow{\nabla}\times\overrightarrow{H} & = & j_{e}+\frac{\partial\overrightarrow{D}}{\partial t}\label{eq:4}\end{eqnarray}

where $\rho_{e}$ and $\rho_{m}$ are respectively the electric and
magnetic charge densities while $\overrightarrow{j_{e}}$ and $\overrightarrow{j_{m}}$
are the corresponding current densities, $\overrightarrow{D}$ is
electric induction vector, $\overrightarrow{E}$ is electric field,
$\overrightarrow{B}$ is magnetic induction vector and $\overrightarrow{H}$
is magnetic field. Here we assume the homogeneous (isotropic) medium
with the following definitions \cite{key-15},

\begin{eqnarray}
\overrightarrow{D} & = & \epsilon\overrightarrow{E}(\epsilon=\epsilon_{0}\epsilon_{r})\label{eq:5}\end{eqnarray}
and

\begin{eqnarray}
\overrightarrow{B} & = & \mu\overrightarrow{H}(\mu=\mu_{0}\mu_{r})\label{eq:6}\end{eqnarray}
where $\epsilon_{o}$ the free space permitivity, $\mu_{o}$ is the
permeability of free space and $\epsilon_{r}$ and $\mu_{r}$ are
defined respectively as relative permitivity and permeability in electric
and magnetic fields. On using equations (\ref{eq:5}) and (\ref{eq:6}),
equations ( \ref{eq:1}- \ref{eq:4}) take the following differential
form,

\begin{eqnarray}
\overrightarrow{\nabla}\cdot\overrightarrow{E} & = & \frac{\rho_{e}}{\epsilon}\label{eq:7}\\
\overrightarrow{\nabla}\cdot\overrightarrow{B} & = & \mu\rho_{m}\label{eq:8}\\
\overrightarrow{\nabla}\times\overrightarrow{E} & =- & \frac{\overrightarrow{j_{m}}}{\epsilon}-\frac{\partial\overrightarrow{B}}{\partial t}\label{eq:9}\\
\overrightarrow{\nabla}\times\overrightarrow{B} & = & \mu\overrightarrow{j_{e}}+\frac{1}{v^{2}}\frac{\partial\overrightarrow{E}}{\partial t}.\label{eq:10}\end{eqnarray}
Differential equations (\ref{eq:7},\ref{eq:8},\ref{eq:9},\ref{eq:10})
are referred as the generalised field equations of dyons in homogeneous
medium and the electric and magnetic fields are corresponding called
generalised electromagnetic fields of dyons. These electric and magnetic
fields of dyons are expressed in following differential form in homogeneous
medium in terms of two potentials \cite{key-11}as,

\begin{eqnarray}
\overrightarrow{E} & = & -\overrightarrow{\nabla}\phi_{e}-\frac{\partial\overrightarrow{C}}{\partial t}-\overrightarrow{\nabla}\times\overrightarrow{D}\label{eq:11}\\
\overrightarrow{B} & = & -\overrightarrow{\nabla}\phi_{m}-\frac{1}{v^{2}}\frac{\partial\overrightarrow{D}}{\partial t}+\overrightarrow{\nabla}\times\overrightarrow{C}\label{eq:12}\end{eqnarray}
where $\{ C^{\mu}\}=\{\phi_{e},v\,\overrightarrow{C}\}$ and $\{ D^{\mu}\}=\{ v\phi_{m},\overrightarrow{D}\}$
are the two four-potentials associated with electric and magnetic
charges. Substituting the values of $\overrightarrow{E}$ and $\overrightarrow{B}$
in (\ref{eq:9}) and (\ref{eq:10}), we get the following sets of
wave equation for dyons in isotropic medium

\begin{eqnarray}
\frac{1}{v^{2}}\frac{\partial^{2}\phi_{e}}{\partial t^{2}}-\nabla^{2}\phi_{e} & =\square\phi_{e} & =\frac{\rho_{e}}{\epsilon};\label{eq:13}\\
\frac{1}{v^{2}}\frac{\partial^{2}\overrightarrow{C}}{\partial t^{2}}-\nabla^{2}\overrightarrow{C} & =\square\overrightarrow{C} & =\mu_{o}\overrightarrow{j_{e}};\label{eq:14}\end{eqnarray}
along with \begin{eqnarray}
\frac{1}{v^{2}}\frac{\partial^{2}\phi_{m}}{\partial t^{2}}-\nabla^{2}\phi_{m} & =\square\phi_{m} & =\mu\rho_{m};\label{eq:15}\\
\frac{1}{v^{2}}\frac{\partial^{2}\overrightarrow{D}}{\partial t^{2}}-\nabla^{2}\overrightarrow{D} & =\square\overrightarrow{D} & =\frac{\overrightarrow{j_{m}}}{\epsilon};\label{eq:16}\end{eqnarray}
where we have imposed the following subsidiary conditions

\begin{eqnarray}
\overrightarrow{\nabla}\cdot\overrightarrow{C}+\frac{1}{v^{2}}\frac{\partial\phi_{e}}{\partial t} & = & 0\label{eq:17}\end{eqnarray}
and

\begin{eqnarray}
\overrightarrow{\nabla}\cdot\overrightarrow{D}+\frac{1}{v^{2}}\frac{\partial\phi_{m}}{\partial t} & = & 0\label{eq:18}\end{eqnarray}
and used the relations $\mu\epsilon=\frac{1}{v^{2}}$or $v=\frac{1}{\sqrt{\mu\epsilon}}=\frac{c}{\sqrt{\mu_{r}\epsilon_{r}}}$
with $c=\frac{1}{\sqrt{\mu_{0}\epsilon_{0}}}$ is the velocity of
light in free space (vacuum) and $v$ is considered as the speed of
electromagnetic wave in homogeneous (isotropic) medium. As such, we
may write the following tensorial representation of generalised Maxwell's
-Dirac (GDM) equations of dyons in homogeneous medium i.e.

\begin{eqnarray}
F_{\mu\nu,\nu} & = & j_{\mu}^{e}\label{eq:19}\\
F_{\mu\nu,\nu}^{d} & = & j_{\mu}^{m}\label{eq:20}\end{eqnarray}
where we have used the definition \cite{key-11} of $F_{\mu\nu}$and
$F_{\mu\nu}^{d}$as

\begin{eqnarray}
F_{\mu\nu} & = & E_{\mu\nu}-H_{\mu\nu}^{d}\label{eq:21}\\
F_{\mu\nu}^{d} & = & H_{\mu\nu}+E_{\mu\nu}^{d}\label{eq:22}\\
E_{\mu\nu} & = & C_{\mu,\nu}-C_{\nu,\mu}\label{eq:23}\\
H_{\mu\nu} & = & D_{\mu,\nu}-D_{\nu,\mu}\label{eq:24}\\
E_{\mu\nu}^{d} & = & \frac{1}{2}\varepsilon_{\mu\nu\rho\sigma}E^{\rho\sigma}\label{eq:25}\\
H_{\mu\nu}^{d} & = & \frac{1}{2}\epsilon_{\mu\nu\rho\sigma}H^{\rho\sigma}.\label{eq:26}\end{eqnarray}

Generalized electromagnetic fields of dyons in homogeneous medium
are thus directly be obtained from field tensor $F_{\mu\nu}$and $F_{\mu\nu}^{d}$as

\begin{eqnarray}
F_{i4} & = & ivE_{i}\label{eq:27}\\
F_{ij} & = & \varepsilon_{ijk}B^{k}\label{eq:28}\\
F_{i4}^{d} & = & ivB_{i}\label{eq:29}\\
F_{ij}^{d} & = & \varepsilon_{ijk}E^{k}.\label{eq:30}\end{eqnarray}
Equations (\ref{eq:11}) and (\ref{eq:12}) are symmetrically invariant
under the following transformations

\begin{eqnarray}
\overrightarrow{E} & \rightarrow & v\overrightarrow{B};\label{eq:31}\\
\overrightarrow{B} & \rightarrow & -\frac{E}{v};\label{eq:32}\\
\overrightarrow{C} & \rightarrow & \frac{\overrightarrow{D}}{v};\label{eq:33}\\
\overrightarrow{D} & \rightarrow & -v\overrightarrow{C};\label{eq:34}\\
\phi_{e} & \rightarrow & v\phi_{m};\label{eq:35}\\
\phi_{m} & \rightarrow & -\frac{\phi_{e}}{v};\label{eq:36}\end{eqnarray}

\begin{eqnarray}
\overrightarrow{j_{e}} & \rightarrow & v\overrightarrow{j_{m}};\label{eq:37}\\
\overrightarrow{j_{m}} & \rightarrow & -\frac{\overrightarrow{j_{e}}}{v};\label{eq:38}\\
\rho_{e} & \rightarrow & \frac{\rho_{m}}{v};\label{eq:39}\\
\rho_{m} & \rightarrow & -v\rho_{e};\label{eq:40}\\
F_{\mu\nu} & \rightarrow & vF_{\mu\nu}^{d};\label{eq:41}\\
F_{\mu\nu}^{d} & \rightarrow & -\frac{F_{\mu\nu}}{v}.\label{eq:42}\end{eqnarray}
Equations(\ref{eq:19}) and (\ref{eq:20}) are also invariant under
the generalized continuous linear transformations \cite{key-16}

\begin{eqnarray}
\overrightarrow{E} & = & \overrightarrow{E}\,\cos\theta+\overrightarrow{B}\, v\,\sin\theta\label{eq:43}\\
\overrightarrow{B}v & = & -\overrightarrow{E}\,\sin\theta+\overrightarrow{B}\, v\,\cos\theta.\label{eq:44}\end{eqnarray}
which reduces to equations (\ref{eq:31} and \ref{eq:32} ) for $\theta=\frac{\pi}{2}$
and thus recalled as duality transformations. Similarly equations
( \ref{eq:33} to \ref{eq:42} ) are also expressed as duality transformations
between electric and magnetic constituents of dyons. As such the GDM
equations given by equations ( \ref{eq:7} to \ref{eq:10} ) are thus
referred as manifestly covariant and dual invariant field equations
of dyons moving in isotropic homogeneus medim. Defining the complex
vector field $\overrightarrow{\psi}$in the following form,

\begin{eqnarray}
\overrightarrow{\psi} & = & \overrightarrow{E}-i\, v\,\overrightarrow{B}\label{eq:45}\end{eqnarray}
and using equations (\ref{eq:11},\ref{eq:12} and \ref{eq:45}) we
get the following relations between generalized field $\overrightarrow{\psi}$
and the components of complex four-potential as 

\begin{eqnarray}
\overrightarrow{\psi} & = & -\frac{\partial\overrightarrow{V}}{\partial t}-\overrightarrow{\nabla}\phi-i\, v\,(\overrightarrow{\nabla}\times\overrightarrow{V)}\label{eq:46}\end{eqnarray}
where $\{ V_{\mu}\}$ is the generalized four-potential of dyons in
homogeneous medium and defined as

\begin{eqnarray}
V_{\mu} & = & \{\phi,\overrightarrow{V}\}\label{eq:47}\end{eqnarray}
i.e. \begin{eqnarray}
\phi & = & \phi_{e}-i\, v\phi_{m}\label{eq:48}\end{eqnarray}
and 

\begin{eqnarray}
\overrightarrow{V} & = & \overrightarrow{C}-i\,\frac{\overrightarrow{D}}{v}.\label{eq:49}\end{eqnarray}
Maxwell's field equation (\ref{eq:7},\ref{eq:8},\ref{eq:9},and
\ref{eq:10}) may then be written in terms of generalized field $\overrightarrow{\psi}$
as

\begin{eqnarray}
\overrightarrow{\nabla}\cdot\overrightarrow{\psi} & = & \frac{\rho}{\epsilon};\label{eq:50}\\
\overrightarrow{\nabla}\times\overrightarrow{\psi} & = & -iv(\mu\overrightarrow{j}+\frac{1}{v^{2}}\frac{\partial\overrightarrow{\psi}}{\partial t});\label{eq:51}\end{eqnarray}
where $\rho$and $\overrightarrow{j}$ are the generalized charge
and current source densities of dyons in homogeneous medium described
as 

\begin{eqnarray}
\rho & = & \rho_{e}-i\,\frac{\rho_{m}}{v};\label{eq:52}\\
\overrightarrow{j} & = & \overrightarrow{j_{e}}-i\, v\overrightarrow{\, j_{m}}.\label{eq:53}\end{eqnarray}
Taking the curl of equation (\ref{eq:51}) and using equation (\ref{eq:50})
we obtain the new parameter (called $\overrightarrow{S}$) expressed
in the following form in terms of source densities i.e.

\begin{eqnarray}
\overrightarrow{S} & =\square\overrightarrow{\psi} & =-\mu\frac{\partial\overrightarrow{j}}{\partial t}-\frac{1}{\epsilon}\overrightarrow{\nabla}\rho-iv\mu(\overrightarrow{\nabla}\times\overrightarrow{j})\label{eq:54}\end{eqnarray}
where $\square$is the D'Alembertian operator and defined as

\begin{eqnarray}
\square & =\frac{1}{v^{2}}\frac{\partial^{2}}{\partial t^{2}}-\nabla^{2} & =\frac{1}{v^{2}}\frac{\partial^{2}}{\partial t^{2}}-\frac{\partial^{2}}{\partial x^{2}}-\frac{\partial^{2}}{\partial y^{2}}-\frac{\partial^{2}}{\partial z^{2}}.\label{eq:55}\end{eqnarray}
Maxwell's-Dirac equation (\ref{eq:7},\ref{eq:8},\ref{eq:9},and
\ref{eq:10}) may now be expressed in the following manner to establish
the relation between generalized potential and current components
of dyons i.e.

\begin{eqnarray}
\square\phi & = & v\mu\rho;\label{eq:56}\\
\square\overrightarrow{V} & = & \mu\overrightarrow{j}.\label{eq:57}\end{eqnarray}
Defining the generalized field tensor of dyon as

\begin{eqnarray}
G_{\mu\nu} & = & F_{\mu\nu}-i\, v\, F_{\mu\nu}^{d}\label{eq:58}\end{eqnarray}
One can directly obtain the following generalized field equation of
dyon in homogeneous (isotropic) medium i.e.

\begin{eqnarray}
G_{\mu\nu,\nu} & = & j_{\mu};\label{eq:59}\\
G_{\mu\nu,\nu}^{d} & = & 0.\label{eq:60}\end{eqnarray}
The suitable manifestly co variant Lagrangian density, which yields
the field equations (\ref{eq:1},\ref{eq:2},\ref{eq:3},\ref{eq:4})
under the variation of field parameters i.e. potential only without
changing the trajectory of particle may be written as follows,

\begin{eqnarray}
L & = & -m_{0}-\frac{1}{4}G_{\mu\nu}G_{\mu\nu}^{*}+V_{\mu}^{*}j_{\mu}\label{eq:61}\end{eqnarray}
where $m_{0}$is the rest mass of particle and {*} denotes the complex
conjugate. Lagrangian density given by equation (\ref{eq:61}) directly
follows the following manifestly covariant and dual invariant form
of Lorentz four-force equation of motion for dyons in homogeneous
(isotropic) medium as

\begin{eqnarray}
f_{\mu} & =m_{0}\ddot{x_{\mu}} & =Re\, q*(G_{\mu\nu}u^{\nu})\label{eq:62}\end{eqnarray}
where $Re$ denotes the real part, $\ddot{x_{\mu}}$is the four-acceleration
and $\left\{ u^{\nu}\right\} $ is the four-velocity of the particle
and $q$ is the generalized charge of dyon defined as follows in isotropic
medium as,

\begin{eqnarray}
q & = & e-i\, v\, g\label{eq:63}\end{eqnarray}
Equations (\ref{eq:19}),(\ref{eq:20}) and (\ref{eq:59},\ref{eq:60})
are invariant under duality transformations,

\begin{eqnarray}
(F,v\, F^{d}) & = & (F\cos\theta+v\, F^{d}\sin\theta,-F\sin\theta+v\, F^{d}\cos\theta)\label{eq:64}\\
(j_{\mu},k_{\mu}) & = & (j_{\mu}\cos\theta+k_{\mu}\sin\theta,-j_{\mu}\sin\theta+k_{\mu}\cos\theta)\label{eq:65}\end{eqnarray}
where

\begin{eqnarray}
\frac{g}{e} & =\frac{B_{\mu}}{A_{\mu}} & =\frac{k_{\mu}}{j_{\mu}}=\frac{F^{d}}{F}=-\tan\theta\label{eq:66}\end{eqnarray}
Hence the the generalized charge of dyon given by equation (\ref{eq:63})
may be written as

\begin{eqnarray}
q & = & |q|\exp[-i\theta].\label{eq:67}\end{eqnarray}
In addition of the dual symmetry the field equations (\ref{eq:59},\ref{eq:60}),
the equation of motion (\ref{eq:62}) and the Lagrangian density (\ref{eq:61})
lead to the following symmetries \cite{key-10};

\begin{enumerate}
\item Invariance under a pure rotation in charge space or its combination
with transformation containing simultaneously space and time reflection
(strong symmetry);
\item A weak symmetry under charge reflection combined with space reflection
or time reflection (not both);
\item A weak symmetry under PT (combined operation of parity and time reversal)
and strong symmetry under CPT ( combined operation of charge conjugation,
parity and time reversal).
\end{enumerate}
Using equation (\ref{eq:67}) the interaction of $i^{th}$dyon with
the field of $j^{th}$dyon in isotropic medium may be written as follows
from the interaction part i.e. the $V_{\mu}^{*}j_{\mu}$part of Lagrangian
density given by equation (\ref{eq:61}), i.e.

\begin{eqnarray}
I_{ij} & = & \frac{C_{\mu}^{(j)}}{e_{j}}q_{j}^{*}q_{i}u_{\mu}^{(i)}\label{eq:68}\end{eqnarray}
where $C_{\mu}^{(j)}$is the electric four-potential describing the
field of $j^{th}$dyon, $e_{j}$is the electric charge and $u_{\mu}^{(i)}$is
the four-velocity of $i^{th}$dyon in the field of $j^{th}$dyon.
This equation shows that

\begin{enumerate}
\item Interaction between two dyon is zero when their generalized charges
are orthogonal in their combined charge space.
\item Interaction depends on electric coupling parameter
\end{enumerate}
\begin{eqnarray}
\alpha_{ij} & = & e_{i}e_{j}+v^{2}g_{i}g_{j}\label{eq:69}\end{eqnarray}

under the constancy condition $\frac{e_{i}}{g_{i}}=\frac{e_{j}}{g_{j}}=constant.$

3. Interaction depends on magnetic coupling parameter (i.e. chirality)

\begin{eqnarray}
\mu_{ij} & = & e_{i}g_{j}-e_{j}g_{i}\label{eq:70}\end{eqnarray}

under the condition $\frac{e_{i}}{vg_{j}}=-\frac{vg_{i}}{e_{j}}.$

In the isotropic (homogeneous) medium the dual invariant energy density
of dyon is now expressed as

\begin{eqnarray}
U & = & \frac{1}{2}\epsilon E^{2}+\frac{1}{2\mu}B^{2}.\label{eq:71}\end{eqnarray}
In equation (\ref{eq:1},\ref{eq:2},\ref{eq:3},\ref{eq:4}) and
(\ref{eq:5},\ref{eq:6}) the generalized Maxwell's equations are
considered together with the so called constitutive relations described
in terms of the relations between the induction vector and field vector.
The constitutive relation given by equations (\ref{eq:5},\ref{eq:6})
then describe the rich variety of physical phenomenon representing
the properties and responses of the medium and to the application
of generalized electromagnetic field of dyons. The field equation,
Lagrangian and the equation of motion of dyons described here in isotropic
medium are also considered as Poincare and conformal invariant, but
there is no trivial gauge group of invariance transformations. Dyonic
fields given by equations (\ref{eq:19},\ref{eq:20}),(\ref{eq:56},\ref{eq:57}),(\ref{eq:59},\ref{eq:60})
and (\ref{eq:62}) are manifestly covariant and also invariant under
duality transformations. Here we have described the electromagnetic
characteristic of dyon field equations in terms of the parameters
of the medium $\epsilon$ and $\mu$ which do not change in time.
Thus our results corresponding to the generalized electromagnetic
models of dyons are represented in terms of time-harmonic (monochromatic)
fields in a consistent and unique manner and reproduces the theories
of the dynamics of electric (magnetic) charge in the absence of magnetic
(electric) charges or vice versa.


\begin{thebibliography}{10}
\bibitem{key-1} G. 't Hooft, Nucl. Phys., \textbf{\underbar{B79}}
(1974), 276.

\bibitem{key-2} A. M. Polyakov, JEPT Letter, \textbf{\underbar{20}}
(1974), 194.

\bibitem{key-3} B. Julia and A. Zee, Phys. Rev., \textbf{\underbar{D11}}
(1975), 2227.

\bibitem[4]{key-4} C. Dokos and T. Tomaros, Phys. Rev., \textbf{\underbar{D21}}
(1980), 2940.

\bibitem[5]{key-5} J. Preskill, Annu. Rev. Nucl. Part. Sci., \textbf{\underbar{34}}
(1984), 461.

\bibitem[6]{key-6} C. G. Callen, Phys. Rev., \textbf{\underbar{D25}}
(1982), 2141.

\bibitem[7]{key-7} V. A. Rubakov, Nucl. Phys., \textbf{\underbar{B203}}
(1982), 211.

\bibitem[8]{key-8} S. Mandelstam, Phys. Rev., \textbf{\underbar{D19}}
(1976), 249.

\bibitem[9]{key-9} E. Witten, Phys. Lett. \textbf{\underbar{86B}}
(1979), 283; G. 't Hooft, Nucl. Phys., \textbf{\underbar{B138}} (1978),
1.

\bibitem[10]{key-10} P. S. Bisht, O. P. S. Negi and B. S. Rajput,
Prog.Theor. Phys., \textbf{\underbar{85}} (1991), 151.

\bibitem[11]{key-11} P. S. Bisht, O. P. S. Negi and B. S. Rajput,
Int. J. Theor. Phys., \textbf{\underbar{32}} (1993), 2099.

\bibitem[12]{key-12} V. V. Kravchenko, {}``Applied Quaternionic
Analysis, Research and Exposition in Mathematics'', Heldermann Press,
Germany, \textbf{\underbar{28}} (2003) and reference therein; {}``Quaternionic
equation for electromagnetic fields in inhomogeneous media'', math-ph/0202010;
Sergei M. Grudsky, Kira V. Khmelnytskaya,Vladislav V. Kravchenko,
{}``On a quaternionic Maxwell equation for the time-dependent electromagnetic
field in a chiral media'', math-ph/0309062.

\bibitem[13]{key-13} P. A. M. Dirac, Proc. Royal Society London,
\textbf{\underbar{A133}} (1931), 60.

\bibitem[14]{key-14}W. A. Barker and Frank Graziani, Am. J. Phys.,
\textbf{\underbar{46}} (1978),1111.

\bibitem[15]{key-15} J. A. Straton, Electromagnetic Theory, McGraw
Hill Company, New York, (1941).

\bibitem[16]{key-16} J. A. Mignaco, Brazilian Journal of Phys., \textbf{\underbar{31}}
(2001), 235.
\end{thebibliography}
\end{document}